\documentclass[aps, prl, reprint, superscriptaddress]{revtex4-1}
\usepackage[utf8]{inputenc}
\usepackage{amsmath}
\usepackage{amsfonts}
\usepackage{amssymb}
\usepackage{graphicx}
\usepackage{todo}
\begin{document}
\title{Multiple-quantum transitions and charge-induced decoherence of donor nuclear spins in silicon}

\author{David P. Franke}
\email{david.franke@wsi.tum.de}
\author{Moritz P. D. Pfl\"uger}
\affiliation{Walter Schottky Institut and Physik-Department, Technische Universit\"at M\"unchen, Am Coulombwall 4, 
	85748 Garching, Germany}
\author{Kohei M. Itoh}
\affiliation{School of Fundamental Science and Technology, Keio University, 3-14-1 Hiyoshi, 
	Kohoku-ku, Yokohama 223-8522, Japan}
\author{Martin S. Brandt}
\affiliation{Walter Schottky Institut and Physik-Department, Technische Universit\"at M\"unchen, Am Coulombwall 4, 
	85748 Garching, Germany}

\begin{abstract}
	We study single- and multi-quantum transitions of the nuclear spins of ionized arsenic donors in silicon and find quadrupolar effects on the coherence times, which we link to fluctuating electrical field gradients present after the application of light and bias voltage pulses. To determine the coherence times of superpositions of all orders in the 4-dimensional Hilbert space, we use a phase-cycling technique and find that, when electrical effects were allowed to decay, these times scale as expected for a field-like decoherence mechanism such as the interaction with surrounding $^{29}$Si nuclear spins.
\end{abstract}

\maketitle
Aiming for the realization of a scalable quantum technology, the electron and nuclear spins of phosphorus donors in silicon have been studied extensively \cite{kane_silicon-based_1998,hollenberg_charge-based_2004,stegner_electrical_2006,morello_single-shot_2010,tyryshkin_electron_2012,steger_quantum_2012,saeedi_room-temperature_2013, buch_spin_2013}. Due to different interaction strengths with their surroundings, they form a powerful combination of a fast, but more volatile electron spin and a slower, but very coherent nuclear spin qubit \cite{morton_solid-state_2008, muhonen_storing_2014, freer_single-atom_2016}. This nuclear spin is $I=1/2$ for phosphorus, but systems with a higher nuclear spin can be realized by simply replacing phosphorus by the other hydrogenic donors As ($I=3/2$) \cite{lo_stark_2014,franke_interaction_2015}, Sb (5/2 and 7/2) \cite{bradbury_stark_2006, salvail_optically_2015}, and Bi (9/2) \cite{wolfowicz_atomic_2013, mortemousque_hyperfine_2014, bienfait_reaching_2016}. Several advantages of the $d$-dimensional Hilbert spaces of such systems, sometimes called qudits, have been proposed, such as the realization of simpler and more efficient gates \cite{ralph_efficient_2007, lanyon_simplifying_2009} or more secure quantum cryptography \cite{cerf_security_2002}. In addition, one higher order system can replace several qubits, simplifying their physical implementation \cite{bartlett_quantum_2002}. These concepts usually require coherent superpositions of higher orders, which show specific interactions with their surroundings that can be reflected in the observed coherence times. In particular, the additional quadrupole interaction with electric field gradients arising from strain \cite{franke_interaction_2015,franke_quadrupolar_2016,pla_strain-induced_2016} or defect states \cite{mortemousque_quadrupole_2015}, has to be considered for heavier dopants. 
In this work, we study first- and higher-order coherences of ionized As donors in silicon. By including light and voltage pulses in the experiment, we are able to link the additional quadrupolar decoherence effect to the electrical environment of the nucleus and show that it vanishes, when the sample is allowed to relax electrically. In addition, we use a phase cycling technique to study superpositions of all orders and find that the coherence times scale inversely proportional to the coherence order, as expected for a field-like decoherence mechanism such as the interaction with surrounding $^{29}$Si nuclear spins. 

\begin{figure}
	\centering
	\includegraphics[width=\linewidth]{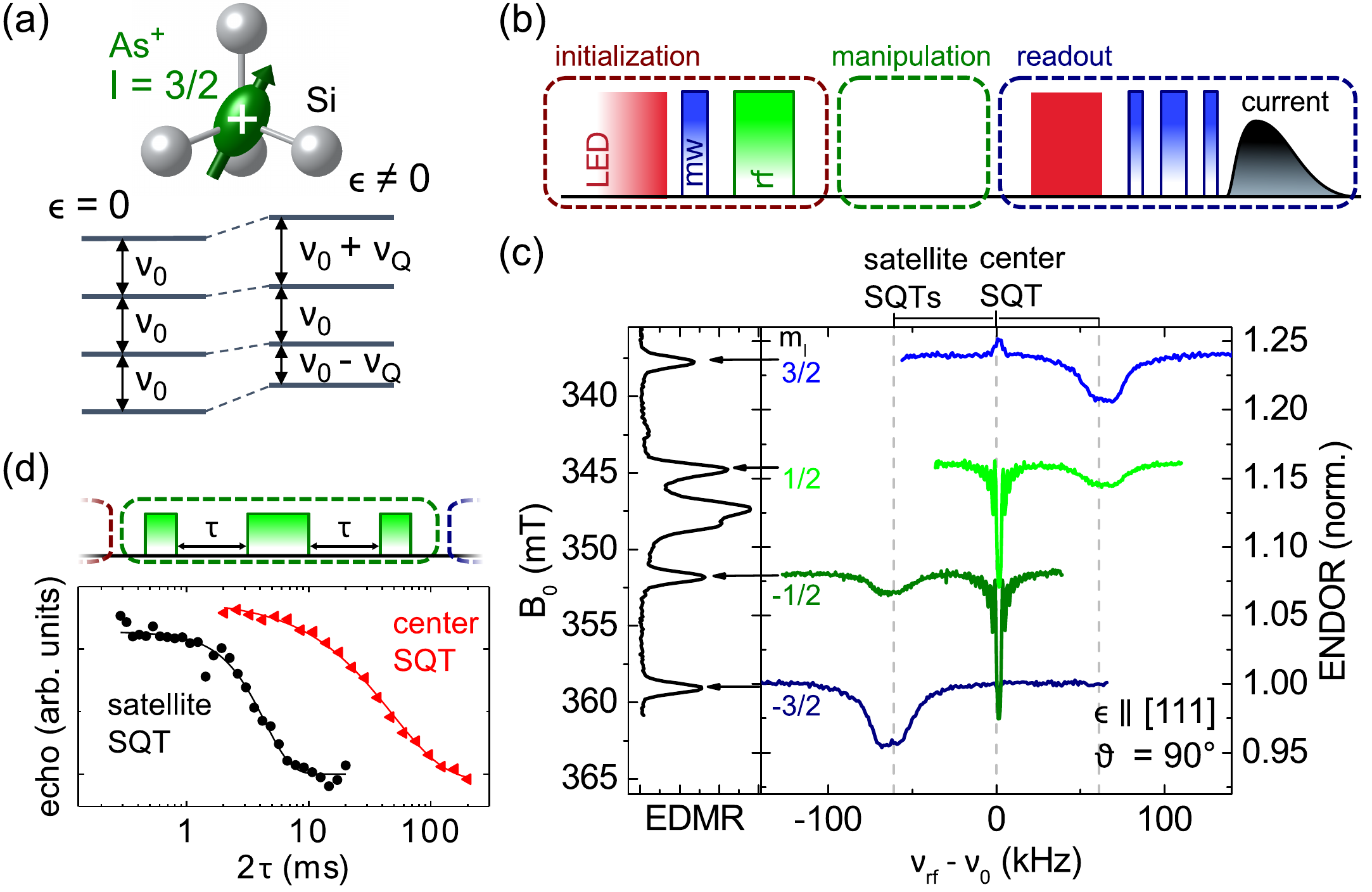}
	\caption{
		(a) Energy levels of the nuclear spin of ionized donors As$^ +$ in Si with and without a uniaxial strain $\epsilon$.
		(b) Schematic representation of the ENDOR pulse sequence used in this work. An example for a manipulation pulse sequence is given on top of (d).
		(c) EDMR and ENDOR spectra of a Si:As sample, where strain is applied along the [111] crystal axis.
		(d) Echo decays for two SQTs of the As$ ^+$ nuclear spin. The solid lines represent fits with (super-)exponential functions (see text).		
		}
	\label{fig:scheme}
\end{figure}

The Hamiltonian $\mathcal{H}$ characterizing ionized arsenic donors with nuclear spin $\mathbf{I}=\{I_x,I_y,I_z\}$ in a magnetic field $B_z$ can be written as
\begin{align}
	\mathcal{H}/h=-\nu_0I_z+\nu_Q\frac{1}{2}\left( I_z^2-\frac{5}{4}\right)
	\mathrm{ ,}\label{eq:H}
\end{align}
where $\nu_0=\gamma_n B_z$ with the nuclear gyromagnetic ratio $\gamma_n$ and $h$ is Planck's constant. The second term on the right hand side of (\ref{eq:H}) describes the nuclear quadrupole interaction with an effective electric field gradient $V_{33}$, here approximated to first order \cite{man_nmr_2012}, with
\begin{align}
	h\nu_Q=\frac{1}{2}V_{33}eQ\cdot\frac{1}{2}(3\cos^2\vartheta -1)\mathrm{ ,}\label{eq:nuQ}
\end{align}
where $Q$ is the nuclear quadrupole moment, $e$ is the elementary charge, and $\vartheta$ describes the angle between $B_z$ and $V_{33}$. The effect of this interaction on the eigenstates of $\mathcal{H}$ is depicted in Fig.~\ref{fig:scheme} (a). For hydrogenic donors in silicon, the quadrupole interaction in general vanishes due to the cubic symmetry of the crystal. Then, the spin states are determined by the nuclear Zeeman interaction only, which leads to four equally spaced levels with transition frequencies $\nu_0$. However, if a strain $\epsilon$ is applied to the sample and this symmetry is broken, $\nu_Q\neq0$ and the eigenenergies are shifted. While the frequency of the central transition is in first order not changed by the quadratic interaction, the two satellite transitions are shifted by equal amounts but in different directions. 

The coherent superpositions, or coherences, in the four-dimensional Hilbert space can be classified by their coherence order $p=m_I^{(i)}-m_I^{(j)}$, where $m_I^{(i)}$ and $m_I^{(j)}$ are the nuclear spin projections of the superimposed states. Following NMR literature, we will call coherences of order $|p|= 1$, 2, and 3 single, double, and triple quantum transitions (SQTs, DQTs, and TQTs), respectively \cite{ernst_principles_1987}. To characterize them, we will follow two different approaches: First, we will study the SQTs in a strained Si sample, where $\nu_Q$ is large compared to the linewidth of the resonance signal \cite{franke_interaction_2015}. In this case, the resonances corresponding to the three SQTs do not overlap  (cf.~Fig.~\ref{fig:scheme} (a)) and can be excited separately; they can be treated as transitions of effective spin-1/2 systems. 
Second, we will measure coherences of all orders in a sample without strain. While, due to selection rules, generally only transitions with $\Delta m_I=\pm1$ can be addressed directly in NMR experiments, higher order coherences can be created by driving degenerate transitions with non-selective pulses, as will be discussed below.

The experiments were performed on [111]- and [110]-oriented Si samples that were implanted with As$^+$ ions and not annealed, conserving the implantation defects, which provides an efficient pair recombination process for electrically detected magnetic resonance (EDMR) \cite{hoehne_time_2013, franke_spin-dependent_2014-1,franke_spin-dependent_2014}. The [111]-oriented sample was thinned and cemented to a sapphire substrate, which at low temperatures generates a compressive strain due to the different thermal extension coefficients of the materials \cite{franke_interaction_2015}. Nuclear spin transitions and echoes are measured by electrically detected electron nuclear double resonance (ENDOR) which is described in detail in Refs.~\onlinecite{dreher_nuclear_2012, hoehne_submillisecond_2015, franke_interaction_2015} and will be reviewed briefly below. The experiments were performed in a Bruker flexline resonator for pulsed ENDOR in a He-flow cryostat at 8 K, illumination was provided by a red light-emitting diode. The samples were biased with typically 8 V; for pulsed voltage experiments, an additional opto-isolator was used for switching.

The basic light, microwave (mw), and radio frequency (rf) pulse sequence used in this study is shown schematically in Fig.~\ref{fig:scheme} (b). Using resonant mw pulses, the selective ionization of donors in a certain nuclear spin state is achieved, leading to a high net polarization of the nuclear spin of the ionized donors. This transient polarization can be driven by rf pulses and is transferred into a stable polarization of all nuclear spins by a $\pi$-pulse (initialization). After all spin pairs have recombined and all donors are in the ionized charge state, the nuclear spin system can be controlled by additional rf pulses, such as echo sequences (manipulation). At the end of the sequence, the sample is illuminated to create conduction band carriers that are trapped by the donors. The nuclear spin state, which is conserved during this capture, is detected via application of an electron spin echo and the evaluation of the current transient after the last mw pulse.

We first study the coherence times $T_2$ of the satellite and center SQTs on the strained sample. In Fig.~\ref{fig:scheme} (c), ENDOR spectra recorded at the four different electron spin resonance fields are shown as a function of the radio frequency $\nu_\mathrm{rf}$. Each field corresponds to the initialization into one nuclear spin state $m_I$, and in each spectrum, the transitions to the neighboring spin states are detected (SQTs), which is why two lines are observed in the spectra for initialization into $m_I=\pm1/2$. As expected for $\epsilon\neq 0$, the satellite SQTs (sSQTs) are shifted by equal amounts, but in different directions $\pm\nu_Q$, while the center SQT (cSQT) is not shifted with respect to $\nu_0$.
An rf Hahn echo sequence with a final projection pulse $\pi/2-\tau-\pi-\tau-\pi/2$ is then inserted into the manipulation part of the pulse sequence and used to determine the coherence times $T_2$ of the center and satellite SQTs. The decays of the echo amplitudes are measured as a function of the evolution time $2\tau$, as shown in Fig.~\ref{fig:scheme} (d). The decays are well described by (super-)exponential functions $\exp(-t/T_2)^\alpha$, with \mbox{$T_2^\mathrm{sSQT}=4.4\pm 1$ ms} ($\alpha=2$) and \mbox{$T_2^\mathrm{cSQT}=48.3\pm 3$ ms} ($\alpha=1$). Since the satellite SQTs, in contrast to the central SQT, are influenced by $\nu_Q$, these results suggest that an additional decoherence mechanism is introduced by the quadrupole interaction.

\begin{figure}
	\centering
	\includegraphics[width=\linewidth]{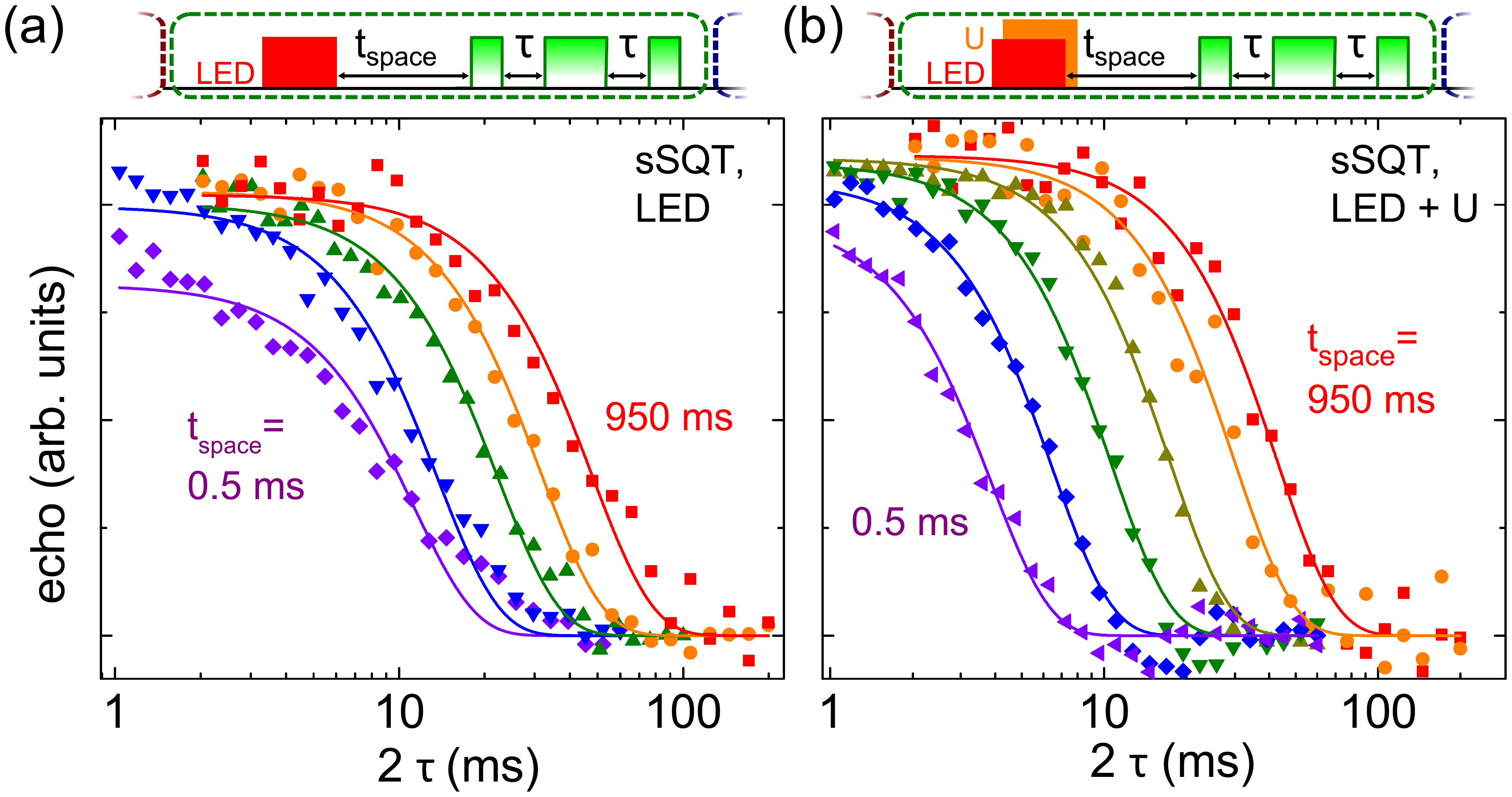}
	\caption{
		Echo decays of a satellite SQT (a) for a 500-$\mu$s-long light pulse and $t_\mathrm{space}=0.5$, 2.5, 20, 80, and 950 ms and (b) for a 500-$\mu$s-long combined light and bias voltage pulse and $t_\mathrm{space}=0.5$, 2.5, 10, 80, 300, and 950 ms. Solid lines represent fits with super-exponential decays ($\alpha=2$).
	}
	\label{fig:decays_U}
\end{figure}

\begin{figure}
	\centering
	\includegraphics[width=\linewidth]{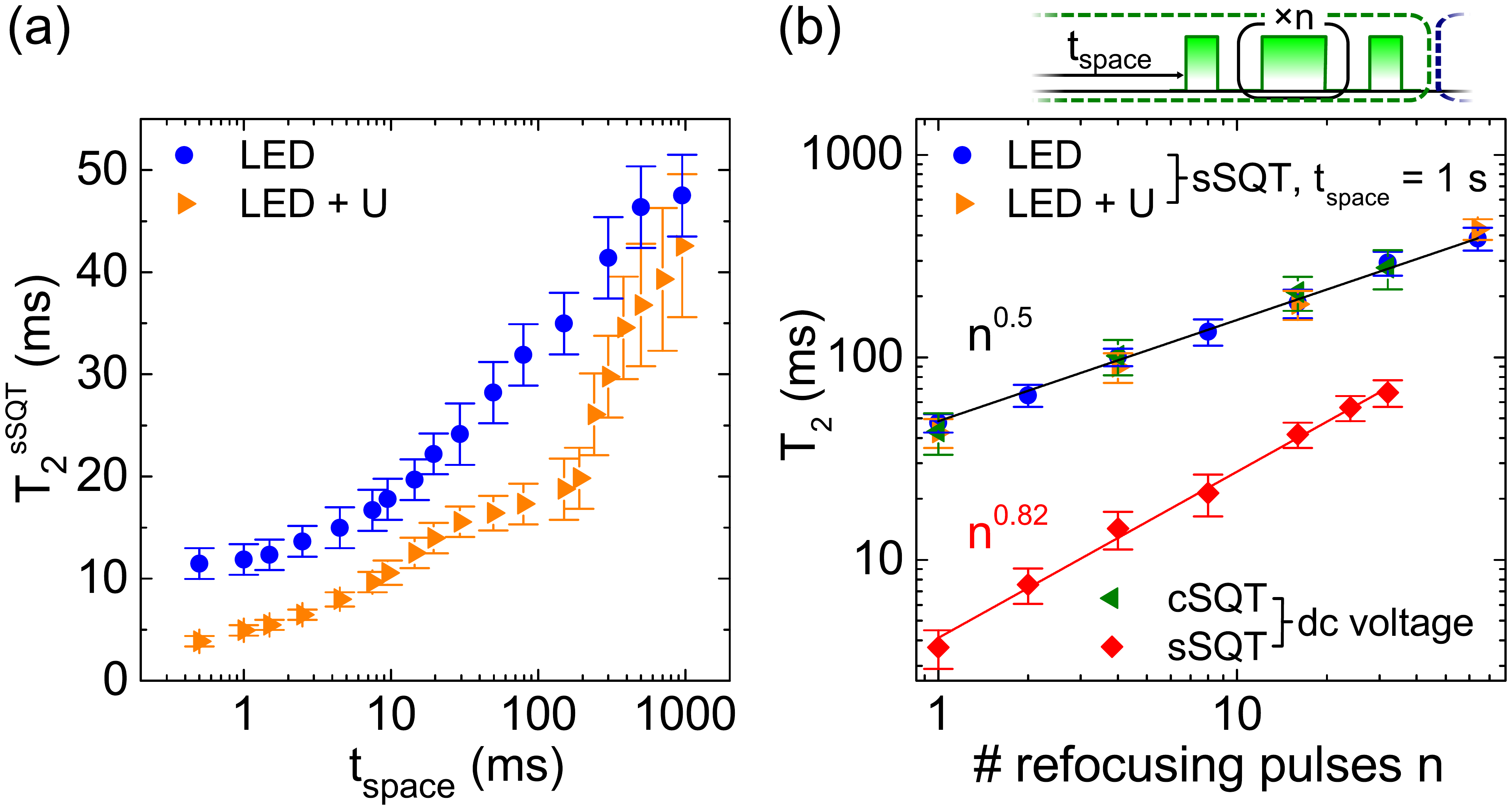}
	\caption{
		(a) Coherence times $T_2^\mathrm{sSQT}$ resulting from the measurements shown in Fig.~\ref{fig:decays_U} (a) and (b).
		(b) Scaling of $T_2$ with the number of refocusing pulses $n$ in a Carr-Purcell sequence. The straight lines are guides to the eye.}
	\label{fig:summary_decays}
\end{figure}

While the quadrupole shifts to the resonance transitions observed in Fig.~\ref{fig:scheme} (c) are the effect of electric field gradients caused by static strain which does not change during the echo experiment and hence will be refocused, additional gradients could potentially be generated by local charges \cite{mortemousque_quadrupole_2015}. Trapping and later recombination of such charges, e.g.~at implantation defects, will lead to changes in these field gradients which would in turn decohere the nuclear spin. To analyze the influence of such effects, we switch from the dc bias voltage applied during the whole sequence to bias pulses. In order to systematically change the electronic environment of the ionized donors in our experiment, we introduce either a light pulse (LED) or a combined light-voltage pulse (LED+U) in the sequence. These pulses are inserted 5 ms after the initialization and a variable time $t_\mathrm{space}$ is added before the nuclear spin echo.
The resulting echo decay traces for light and light-voltage pulses are shown in Fig.~\ref{fig:decays_U} (a) and (b), respectively. A strong influence of $t_\mathrm{space}$ on the decay is observed in both experimental series. The decays can be fitted by super-exponential functions ($\alpha=2$), and the extracted coherence times $T_2$ are plotted as a function of $t_\mathrm{space}$ in Fig.~\ref{fig:summary_decays} (a). Clearly, the coherence time rises systematically with longer $t_\mathrm{space}$, and seems to approach saturation at $T_2\approx 50$ ms for $t_\mathrm{space}>1$ s, where the experiment was discontinued because of to the long measuring times involved. However, the values at 1 s are within error equal for both measurement series and also equal to the $T_2$ of the center SQT, which is not influenced by the implementation of the additional pulses (data not shown).
To verify that the additional decoherence process has indeed ceased after this time, we measure the decays for Carr-Purcell sequences \cite{carr_effects_1954} with different numbers of refocusing pulses $n$, as shown in Fig.~\ref{fig:summary_decays} (b). Within error, the observed scaling laws $T_2\propto n^\beta$ are equal for the satellite SQT for $t_\mathrm{space}= 1$ s (LED and LED+U) and the center SQT, the exponent $\beta = 0.5$  indicates a noise spectrum with a $1/f$ frequency dependence \cite{medford_scaling_2012}.
The application of (dark) bias pulses during the nuclear echo sequence did not have any effect on the coherence time, indicating that the influence of field gradients created directly by the contact structures can be neglected in our experiments. We can therefore conclude that the additional decoherence that acts on the satellite SQTs is connected to the immediate electronic environment of the ionized donors which is influenced by the light and voltage pulses. In particular, we suspect that after the generation of conduction band charge carriers, the trapping and recombination of these charges at defect states lead to fluctuating electric field gradients.
The higher exponent $\beta=0.82$ for quadrupolar decoherence effects observed for the satellite SQT in the dc bias measurements reflects the stronger effect of the decoupling in the case of the relatively slow changes in the charge environment. 

\begin{figure}
	\centering
	\includegraphics[width=\linewidth]{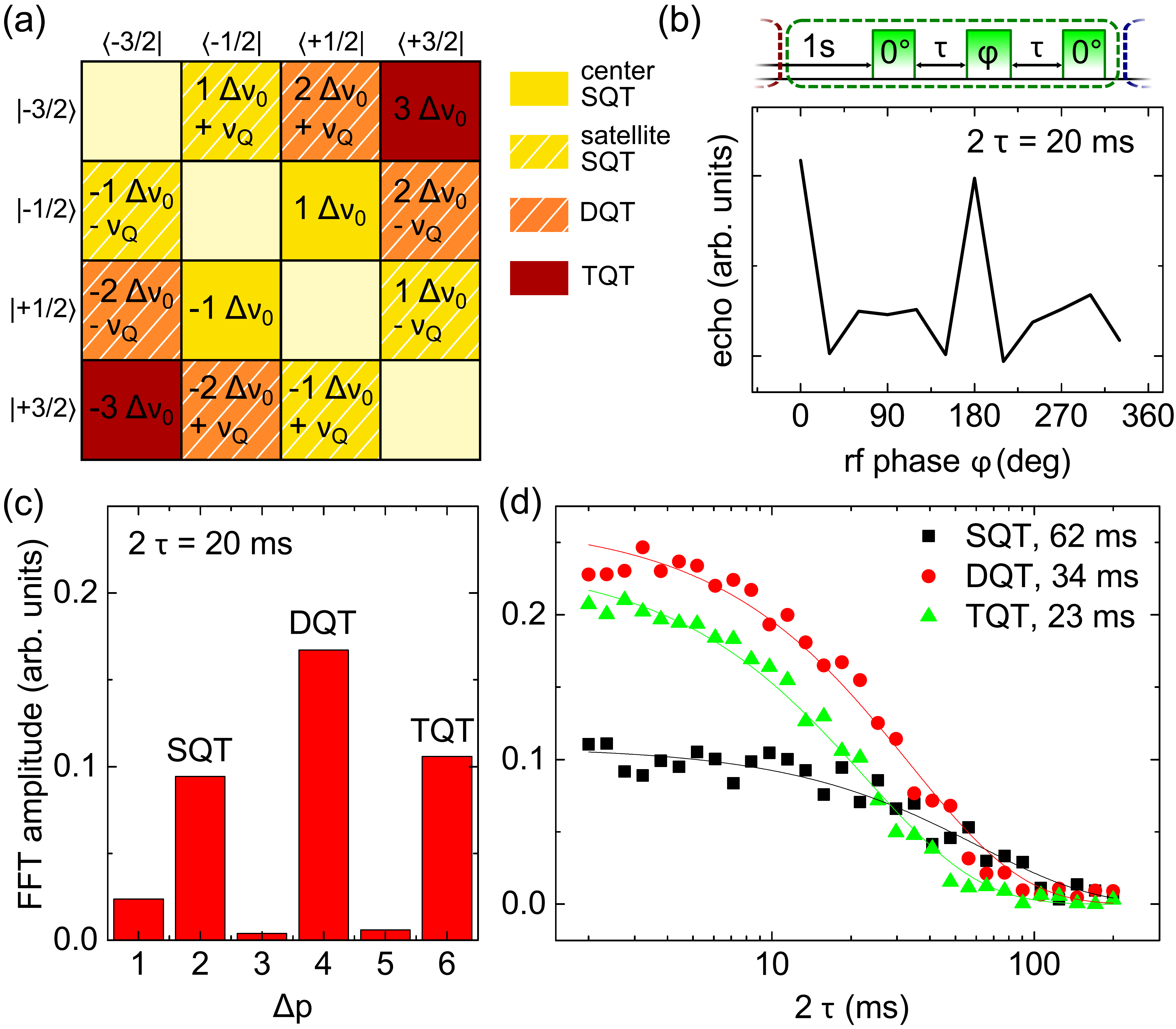}
	\caption{
		(a) Summary of the time evolution frequencies $\Delta\nu_{ij}$ of the density matrix elements.
		(b) Echo amplitude as a function of the phase of the second pulse of the nuclear spin echo sequence.
		(c) Fourier transform, showing the contributions of different coherence transfer paths to the echo signal. For details see text.
		(d) Echo decays for the SQT ($\Delta p=2$), DQT ($\Delta p=4$) and TQT ($\Delta p=6$) signals for spacing $t_\mathrm{space}=1$ s. The solid lines are fits with exponential functions.}
	\label{fig:higher_order}
\end{figure}

To study coherences of all orders, we measure the unstrained sample, where the SQTs are degenerate and nonselective rf pulses can be applied. We first consider the density matrix $\rho$ of the nuclear spin ensemble, where the eigenstates are denoted by the diagonal entries and the coherences are given by the off-diagonal entries. 
Since $\mathcal{H}$ is diagonal, it can be written as \mbox{$\mathcal{H}^\mathrm{rot}=\mathcal{H}+h\nu_\mathrm{rf}I_z$} in the rotating frame and the time evolution of the entries $\rho_{ij}$ is given by
\begin{align}
	\rho_{ij}(t)=\rho_{ij}(0)\cdot e^{i2\pi\Delta\nu_{ij}t}\text{ ,}
\end{align}
where $h\Delta\nu_{ij}=\mathcal{H}^\mathrm{rot}_{ii}-\mathcal{H}^\mathrm{rot}_{jj}$. In Fig.~\ref{fig:higher_order} (a), the $\Delta\nu$ are shown in an array plot symbolically representing the density matrix $\rho$. The orders of the corresponding coherences are indicated by colors; in addition, the transitions that are influenced by first order quadrupole interaction are shaded in white. Again, the difference between the satellite and center SQTs becomes clear, as the evolution of the former includes $\nu_Q$. The same is true for the DQTs, while the TQT, like the center SQT, is independent of $\nu_Q$. Furthermore, the evolution of each coherence of order $p$ includes the term $p\cdot\Delta\nu_0$, where $\Delta\nu_0=\nu_0-\nu_\mathrm{rf}$. Note that even in the nominally ``unstrained'' samples measured in this work, a significant distribution of $\nu_Q$ is observed and dominates the linewidth and the dephasing of the relevant transitions \cite{franke_interaction_2015}.

To determine $T_2$ for the different coherences of the As$^+$ nuclear spin, we will again measure nuclear spin echoes. It is therefore instructive to consider the principle of a spin echo in the density matrix formalism. The first pulse in an echo sequence is used to excite coherences from the initial groundstates. They evolve with their respective $\Delta \nu_{ij}$ and during the time $\tau_1$, each spin collects a phase $\Delta\nu_{ij}\cdot\tau_1$. Then, the second (refocusing) pulse is applied, which transfers the coherence from $\rho_{ij}$ to $\rho_{i^\prime j^\prime}$. It will again evolve, and after the time $\tau_2$, each spin has collected a total phase of $\Delta\nu_{ij}\cdot\tau_1+\Delta\nu_{i^\prime j^\prime}\cdot\tau_2$. Hence, if $\Delta\nu_{ij}=-\Delta\nu_{i^\prime j^\prime}$, the total phase vanishes for $\tau_1=\tau_2=\tau$, independently of the actual value of the $\Delta \nu$. Any distribution of $\Delta\nu$ of the ensemble will then refocus, which is the well-known echo effect. Since the propagator is Hermitian, $\Delta\nu_{ij}=-\Delta\nu_{ji}$, and as can be seen with help of Fig.~\ref{fig:higher_order} (a), the refocusing condition only holds for exactly these pairs. 
A $\pi$-pulse, used as refocusing pulse in the conventional echo sequence, leads to $m_I\rightarrow-m_I$ for all states, which is equivalent to flipping the entries of the density matrix across its center. Comparing the corresponding $\nu_{ij}$, it becomes clear that the refocusing condition is in this case only fulfilled for the center SQT and the TQT, while satellite SQT and DQT are not refocused because of the distribution of the $\nu_Q$. 
To achieve echoes also from these coherences, we use an echo sequence consisting of $3$ pulses with lengths of 45, 50, and 45 $\mu$s, each corresponding to a nutation angle of $\sim 2\pi/3$ \cite{solomon_multiple_1958, abe_spin_1966}.
Since in such a sequence several different echoes are observed simultaneously, it is necessary to separate the signals corresponding to the refocusing of different coherences. To this end, we use a phase cycle of the phase $\varphi$ of the refocusing pulse. This adds a shift $\varphi\Delta p$ to the phase of the coherence that is transferred from $\rho_{ij}$ to $\rho_{i^\prime j^\prime}$, where $\Delta p = p-p^\prime$ is the difference in coherence order of $\rho_{ij}$ and $\rho_{i^\prime j^\prime}$. Therefore, the detected echo signal will be modified as a function of $\varphi\Delta p$ and the different components can be separated by a Fourier transform, equivalent to coherence pathway selection in NMR \cite{ernst_principles_1987}.

We apply the above-mentioned echo sequence after initialization on $m_I=-1/2$. The measured echo amplitude as a function of $\varphi$ is shown examplarily in Fig.~\ref{fig:higher_order} (b), its Fourier transform is given in Fig.~\ref{fig:higher_order} (c).  The largest part of the echo signal is due to a $\Delta p=4$ coherence transfer (corresponding to the DQT echo), additional components result from $\Delta p=6$ (TQT) and $\Delta p= 2$ (SQT). For short $\tau$, significant signals are observed also for $\Delta p=1$ and 3. Since $\Delta\nu_0$ cannot be refocused in such a coherence path, these contributions vanish with $\tau>T_2^\ast$ which is $\sim 5$ ms in our experiments.
To measure the different coherence times, we change $\tau=\tau_1=\tau_2$ and measure the decays of the three coherences. These experiments are performed with a waiting time $t_\mathrm{space}=1$ s to minimize effects of charges created by the light and bias pulses during the readout part of the pulse sequence.
The resulting traces are shown in Fig.~\ref{fig:higher_order}(d) and can be fitted with exponential functions. The extracted coherence times $T_2$ are $62\pm 10$, $34\pm 5$ and $23\pm 5$ ms for SQT, DQT and TQT, respectively. While we cannot differentiate between the different SQTs, the observed $T_2^\mathrm{SQT}$ is in agreement with $T_2^\mathrm{sSQT}$ and $T_2^\mathrm{cSQT}$ we found in the strained sample above, confirming that quadrupolar decoherence effects are negligible for the chosen $t_\mathrm{space}$.

Any decoherence mechanism that can be treated as an effective magnetic field will introduce a fluctuation of $\Delta\nu_0$. Since this is reflected in the time evolution as $p\Delta\nu$, we expect the effect of such disturbances on $T_2$ to be anti-proportional to $p$. Accordingly, the ratios \mbox{$T_2^\mathrm{SQT}: T_2^\mathrm{DQT}: T_2^\mathrm{TQT} = 6:3:2$} should be observed if the limiting process is an effective field interaction.
This is in good agreement with experiment, suggesting that a field-like interaction, such as the dipolar interaction with environmental $^{29}$Si spins \cite{itoh_isotope_2014, petersen_nuclear_2016} or fluctuations of the external magnetic field, is responsible for the decay of all coherences in the As$^+$ nuclear spin system. Since we have observed significantly longer coherence times in purified $^{28}$Si samples using the same magnetic field setup \cite{franke_electron_2016}, we can conclude that interactions with the $^{29}$Si nuclear spin bath limit the coherence in our samples.

In summary, we have characterized coherences of first and higher orders of the nuclear spin of ionized As donors in silicon with and without strain. We have found that, after the application of light and light-voltage pulses, quadrupolar effects can limit the coherence some transitions and that these influences are minimized when allowing for longer waiting times during the pulse sequence. Such decoherence could be relevant in any nuclear spins system with $I>1/2$, and can be avoided by cosing optimal working points in the magnetic field and strain, equivalent to clock transitions with respect to magnetic field \cite{wolfowicz_atomic_2013} or hyperfine interaction \cite{mortemousque_hyperfine_2014}.
We have also measured echoes connected to higher-order coherences using a phase-cycling technique, exploring the full potential of the four-dimensional Hilbert space. The scaling of the observed coherence times suggest that $T_2$ in our samples is limited by the interaction with the $^{29}$Si nuclear spin bath and the applied technique could be of interest for measurements on donor nuclear spins $I>1/2$ in isotopically purified $^{28}$Si, where this influence is minimized and the limits of the resulting very long coherence times can be explored.
For applications with higher-order quantum systems, usually a more precise control of the higher order coherences will be needed. Pulse shaping and optimal control pulses  \cite{skinner_application_2003, lee_optimal_2008} could possibly enable a deterministic excitation of any of the coherences in lightly strained samples and could also allow for a dynamical decoupling for all coherences.

\begin{acknowledgments}
The authors would like to thank Hans-Werner Becker for the implantation, Manabu Otsuka for sample characterization, and Steffen Glaser for fruitful discussion. The work at TUM was supported financially by DFG via SFB 631 and SPP 1601, the work at Keio by KAKENHI (S) No. 26220602 and JSPS Core-to-Core.
\end{acknowledgments}

\todos

\bibliography{bib4}
\end{document}